\documentclass{ws-rv961x669}
\usepackage{ws-rv-van}     % numbered citation/references (default)
\usepackage{ws-rv-thm}     % comment this line when `amsthm / theorem / ntheorem` package is used
\usepackage{subfigure}     % required only when side-by-side / subfigures are used
\makeindex
\usepackage{mathrsfs,amsmath,graphicx,paralist}
\def\avg#1{\langle#1\rangle}    \def\<{\langle}         \def\>{\rangle}

\def\al{\alpha}                   

\def\sig{\sigma}        \def\del{\delta}        \def\Del{\Delta}
\def\eps{\epsilon}        
\def\up{\uparrow}       \def\down{\downarrow}

  \def\Vq{{\mathbf q}} 
    
  \def\V0{{\mathbf 0}}

  \def\B0{{\mathbf 0}}
\def\Br{{\bf r}}   

 \def\Bp{{\mathbf p}} \def\Bq{{\mathbf q}}

  \def\BQ{{\mathbf Q}}

\def\be{\begin{equation}}       \def\ee{\end{equation}}
\def\bea{\begin{eqnarray}}      \def\eea{\end{eqnarray}}

\begin{document}

\def\chaptertitle{Breached pair superfluidity: a brief review}
%\chapter{Breach pair superfluidity: a brief review}
\begin{center}
  {\bf\large \chaptertitle}
\end{center}
%\vskip 10pt

\author{W. Vincent Liu\footnote{Initial work submitted during sabbatical leave
    from the present address: Department of Physics and Astronomy, University of
    Pittsburgh, Pittsburgh, Pennsylvania 15260, USA. Email: wvliu@pitt.edu.}}

\address{Department of Physics and Shenzhen Institute for Quantum Science and
  Engineering, Southern University of Science and Technology, Shenzhen 518055,
  China and \\
  Wilczek Quantum Center, School of Physics and Astronomy and T. D. Lee Institute, Shanghai Jiao Tong University, Shanghai 200240, China}

% \author{W. Vincent Liu\footnote{Email: wvliu@pitt.edu.}}
% \address{Department of Physics and
% Astronomy, University of Pittsburgh, Pittsburgh, Pennsylvania 15260, USA}

%%\date{\today}

\begin{abstract}
 
  Interior gap superfluidity was introduced together with Frank Wilczek. Later
  on together with our collaborators, we generalized this new possibility of
  superfluidity to a broader concept, breached pair superfluidity. In the
  occasion to celebrate Professor Frank Wilczek's seventieth birthday and his
  productive career in several major areas in physics, I dedicate this note to
  recall the exciting times of developing this idea, the main aspects of the
  proposed phase, and the discussion on its stability condition.

\end{abstract}

% \markboth{Even Page Header}{Odd Page Header} % Customized running heads
\markboth{W. Vincent Liu}{\chaptertitle} % Customized running heads

\body

%\tableofcontents

\section{Introduction}

Modern history of physics has proudly recorded a group of great luminaries
having broadly impacted on the areas of high energy, statistical, and condensed
matter physics. Among those who are still live and active in today's physics,
many would immediately point out Frank Wilczek in this class.  I was lucky to
have had the privilege of working with him---during my postdoctoral times at
MIT---on one of the topics of great interest in such an interdisciplinary
area. It was about exotic superfluidity arising from the effect of mismatched
Fermi surfaces.

The phenomenon of mismatched Fermi surfaces occurs in a number of physical
systems.   It happens in electronic superconductors when an in-plane
magnetic field is applied or when superconductivity and ferromagnetism
coexist.  It should happen in dense quantum chromodynamics (QCD)
matter---arguably the inside of neutron stars or a quark-gluon plasma
realized in the heavy ion collision experiments at CERN \cite{Pasechnik:17rev}.  It
can be artificially tuned to occur, not as a problem but as a new parameter
regime for possible new effects, in cold atomic gases through the mixtures of
different atomic species whose populations are separately controlled and
conserved.

Understanding the Zeeman effect in the electronic superconductors was a key
motivation in the early proposal by Fulde-Ferrell-Larkin-Ovchinnikov (FFLO)
\cite{FFLO:64+65} to consider the instability of Bardeen-Cooper-Schrieffer (BCS
\cite{BCS:57}) phase towards alternative energetically favored states to
accommodate the Fermi level difference due to spin polarization. This class of
states are now known as FFLO phases, in which each Cooper pair carries a finite
center-of-mass momentum whose scale is set by the Fermi surface difference,
i.e. $\BQ \sim 1.2 \del p_F$ by mean field theory.

Frank Wilczek and his collaborators are among the early pioneers to initiate the
field of color
superconductivity~\cite{Alford+Wilczek:98,Rapp+:98,Alford+Wilczek:99,+Wilczek:00,QCD-FFLO:04rmp,Alford+:08rmp}.
The condensed matter concept of FFLO phase was extended to the understanding of
the phase diagram of dense QCD matter, for which crystalline color
superconductivity was
proposed~\cite{+Rajagopal:01,Bowers-Rajagopal:02,Alford+:08rmp}. The study of
quark matter at finite density turns into a new condensed matter physics
problem.  A fundamental change in comparison is the role of the usual Coulomb
interaction between electrons mediated by U(1) electromagnetic gauge bosons
(photons) being replaced by the strong interaction mediated by SU(3) color gauge
bosons known as gluons.  And there are more degrees of freedom and higher
symmetry involved, such as a variety of Cooper pairs with several flavors and
colors present at the appropriate density level (for that matter, also the
energy scale).

In high density QCD, the phenomenon of mismatched Fermi surfaces is guaranteed
to exist by two facts combined: six quark flavors carry two different fractions
of electrical charge and each flavor has its own unique mass, hence different
from each other.  Let us illustrate this by considering an intermediate high
density quark matter with average chemical potential $\mu\sim 400$MeV, which
probably corresponds to the nucleon density in the core of neutron stars.  At
such an energy scale, the charm, top and bottom flavors are not present due to
their individual masses being significantly high above.  The QCD matter then is
made of (approximately) massless up and down quarks and massive strange quarks.
The strange quark has a mass $M_s \sim 300$MeV, comparable to the chemical
potential scale under consideration.  It plays a crucial role, as we see below,
to cause a mismatch in flavor chemical potentials.  Electric neutrality requires
a nonzero density of electrons present, estimated to be $\mu_e \sim 50$MeV.  In
many body physics, $\mu$ and $\mu_e$ are treated as Lagrangian multiplier to
enforce the conserved quantum numbers of quark flavor and electric charge,
respectively.  A simple algebra shows that they are related to the individual
flavor chemical potentials and Fermi momenta as follows:
\begin{equation}
\begin{array}{ll}
  \displaystyle
  \mu_u =\mu -{2\over 3} \mu_e =367 {\rm MeV}\,, \qquad & p_F^u
                                                              =\mu_u= 367 {\rm MeV}\,, \\
 \displaystyle \mu_d =\mu +{1\over 3} \mu_e = 417 {\rm MeV}\,, & p_F^d =\mu_d 
                                                    = 417 {\rm MeV} \,,\\
 \displaystyle  %\textstyle 
  \mu_s =\mu +{1\over 3} \mu_e = 417 {\rm MeV}\,, & p_F^s
                                                    =\sqrt{\mu_s^2-M_s^2} =289
                                                    {\rm MeV}\,.
\end{array}
\end{equation}
The above crude estimate shows that the three flavors would have three different
Fermi surfaces. An immediate consequence is that the color-flavor-locking
superconductivity would be unstable towards crystalline color superconductivity,
the manifestation of FFLO in high density QCD.  This was analyzed in the context
of color superconductivity by Alford, Bowers, and
Rajagopal~\cite{+Rajagopal:01,Bowers-Rajagopal:02} (for reviews, see
~\cite{+Wilczek:00,QCD-FFLO:04rmp,Alford+:08rmp}).  The BCS equivalent in this
context is the color-flavor-locking (CFL) superconductivity with equal
population pairing, proposed for the extreme high density limit where all
flavors would enjoy approximately the same Fermi surfaces with chemical
potential difference negligible at such a high-density scale.  It was discussed
in 1970s and 1980s (see Bailin and Love
\cite{Bailin-Love:79,Bailin-Love:84Rep}) that dense quark matter might become
superfluid.

Frank, together with his collaborators, pioneered in revitalizing the idea of
color superconductivity.~\cite{Alford+Wilczek:98,Rapp+:98,Alford+Wilczek:99,+Wilczek:00,QCD-FFLO:04rmp,Alford+:08rmp}
He also knew well of the FFLO idea from his close collaborators and former
students who extended the FFLO concept to the intermediate density regime of
quark matter to introduce crystalline color superconductivity.  I was
fortunate to be exposed to this interesting interface between condensed matter
and high energy physics, immediately after I went to MIT as a postdoctoral
fellow with Frank in 2001. I quickly learned a great deal of color
superconductivity and QCD matter phase diagram from him, and we realized that
some of the ideas could be quantum simulated by ultra-cold atomic gases.

Progressive developments in ultra-cold gases have revitalized interest in some
basic qualitative questions of quantum many-body theory, because they promise to
make a wide variety of conceptually interesting parameter regimes, which might
previously have seemed academic or excessively special, experimentally
accessible.  Advances in the spatiotemporal control and readout of ultra-cold
quantum gases are rapidly expanding the scientific range of current and
near-future experiments in this growing field.\cite{UCA:nature}.  Emergent tools
of quantum control in AMO physics represent an important bridge to the
exploration of complex many-body problems based on fully understood few-particle
subsystems.  One of the most remarkable is the unprecedented flexibility to
directly control the interaction between the selected atomic internal
states---which play the role of ``spins'' or `` flavors''---through the
technique known as Feshbach resonance
\cite{Ketterle:07rev,Giorgini+Stringari:08rev,Carr+Ye:09rev,Chin-Grimm:10rmp,Hulet_Fermi:12rev}.
It has been demonstrated that interaction between fermionic atoms can be
precisely tuned, by dialing an external magnetic field, from attractive to
repulsive, from weak to infinitely strong. Trapping cold atoms and molecules in
optical lattices brings a whole new set of quantum
manipulations.\cite{Jaksch-Zoller:05rev,Lewenstein:07rev,Bloch:08rmp} Among
them, the band masses and interaction between atoms on the same or neighboring
sites are all becoming experimentally controlled with unprecedented
flexibilities. Novel forms in geometry or (dynamic) Floquet engineering have
been demonstrated and used to explore a wide range of interesting quantum states
of matter, from simulating topological insulators and fractional quantum Hall
effects to exploring antiferromagnetism and orbital superfluidity.
\cite{Tilman:10rev,Stamper-Kurn_spinor:13rmp,Lewenstein:15rpp,Hemmerich_synopsis:16jpb,Li-Liu:16RPP}

\section{Aspects of breached pair superfluidity}
The FFLO actually represents a class of superconducting states, because the
FFLO order parameter in principle can be a superposition of Cooper pairs
condensed at a set of different finite momenta,
\begin{equation}
\Delta (\Br)\equiv
\avg{\psi_\sig(\Br) \psi_{\sig'}(\Br)\epsilon_{\sig\sig'} } \sim \sum_{\{
  \BQ_\al\}}\al e^{i\BQ_\al\cdot \Br}\Delta_\al
\end{equation}
where I assume a fermion model of two spins $\sig=\up, \down$,
$\epsilon_{\sig\sig'} $ is the antisymmetric tensor, and the summation over
$\al$ represents a set of Cooper pair momenta under choice to minimize the
postulated ground state energy.  Hence, it yields a rich phase diagram of
different ordering crystalline structures.  In other words, generic FFLO phases
break both spatial translational and rotational symmetries in addition to the
phase U(1) rotation symmetry.  A FFLO pairing order parameter with a single
momentum component is an exception as it would be still translationally
invariant. The rich phase diagram however poses an experimental challenge to
realize, observe and identify the nature and symmetry of the actual FFLO phase
being realized, in part due to an abundance of competing FFLO phases which
differ in crystalline symmetry.  The energy difference between similar but
symmetrically different FFLO phases often is very small, from variational
calculations. The overall condensation energy saved in FFLO scenario is also
exponentially smaller than that of the corresponding BCS phase if the condition
of Fermi surface match restores while everything else being kept the same.  The
reason is quite easy to understand. Unlike the BCS case, FFLO pairing only takes
place in certain spots of the Fermi surfaces, so the density of states of
fermionic particle states participating in pairing and condensation is
fractional, not a whole shell along the Fermi surface in a 3D setting.  The
energy gap and critical temperature both depend on the density of states of
paired fermions exponentially, if the BCS mean field theory is taken as guidance.

The new experimental regime in cold atomic gases  motivated Frank and me to
think whether it was possible to have some homogeneous phase ---on this regard like the
BCS but not like the FFLO---that can be energetically better than the
non-homogeneous FFLO.    Our initial proposal, as a phenomenological trial
wavefunction for the superfluid ground state was interior gap superfluidity\cite{Liu-Wilczek:03}.
Later on, we realize this represents just one special limit of a more general
possibility\cite{Gubankova+:03,Liu+Zoller:04,Forbes:2004cr}. In the following, I will first review the general state and then
point the interior gap as a special situation of the general scenario.

%{\bf later on exterior gap -> BP}

Let us use an example of model Hamiltonian to introduce the breached pair
superfluid (BP) state\cite{Gubankova+:03,Forbes:2004cr}, 
\begin{equation}
H= \sum_{\Bp\sig} \left[ \frac{\Bp^2}{2m_\sig} - \mu_\sig\right]
{\psi_{\sig\Bp}}^\dag {\psi_{\sig\Bp}} 
 + \sum_{\Bp\Bp^\prime} V_{\Bp\Bp^\prime}(\Bq) \psi^\dag_{\up \Vq+\Bp}
\psi^\dag_{\down \Bq-\Bp} \psi_{\down \Bq-\Bp^\prime}
 \psi_{\up\Vq+\Bp^\prime} \,, 
\end{equation}
where  $\sig=\{\up,\down\}$ (or $\{A, B\}$) is the spin indices, $m_\sig$ are
the masses of fermions, and $\mu_\sig$ are the chemical potentials to enforce
the condition of spin population imbalance, say $N_\up < N_\down$. Unlike 
previous studies, a key additional feature that I believe was first explored by
Frank and me \cite{Liu-Wilczek:03} is the introduction of mass imbalance $m_\up \neq m_\down$.
It was found that the larger the mass ratio is, the better the BP state is
favored. As a concrete example, 
 assume that the spin down species is much heavier than the spin up species
 ($m_\down \gg m_\up$)  without loss of generality. In order for the BP state to
 be energetically favorable, the interaction $V_{\Bp\Bp^\prime}(\Bq) $ has to be
 strongly momentum dependent. (A contact $\del$-like interaction in the real
 space corresponds to a $\Bq$-independent constant  in momentum space by
 Fourier transformation, which does not work.)   Two types of interaction, each
 of special interest, were found to work\cite{Forbes:2004cr}. One is of the
 (modified) BCS type of
 interaction
 \begin{equation}
   V^I_{\Bp\Bp^\prime}(\Bq) =\left\{
     \begin{array}{ll}
       g \ \mbox{(constant)}  \qquad & \mbox{if}\quad  p,p^\prime \in
                                       [p^F_\sig-\lambda/2, \ p^F_\sig+\lambda/2]\,,
       \\
       0\,.
     \end{array} \right.
 \end{equation}
 where $p^F_\sig$ are the Fermi momentum for the two spins ($\sig=\up,\down$)
 and the non-vanishing condition in the above is for momenta sitting within the
 stripes of width $\lambda$ around each of two Fermi surfaces.  Here is a subtle
 point---$\lambda$ behaves like the ultraviolet cutoff, in the sense
 $\lambda \gg \Delta/v_F$ (where $\Del$ is the energy gap due to Cooper pairing), but should be kept smaller than the Fermi surface
 mismatch, $\lambda < \del p^F=p^F_\down-p^F_\up$.  This type of model was used
 in the calculation of our 2003 PRL~\cite{Liu-Wilczek:03} when first introducing
 the concept of interior gap superfluidity.  Such a condition was briefly
 indicated in a sentence put in parentheses under Eq.~(1) in
 Ref. \cite{Liu-Wilczek:03}. The work by Wu and Yip\cite{Wu-Yip:03} was on a qualitatively
 different model--- their interaction is point-like in the position space, which
 transforms into a uniform constant in momentum space with a momentum cut-off
 scale taken to be the largest scale, $\gg \del p^F$. In other words, the type
 of interaction considered in Wu-Yip paper is not what used in our model
 calculation. This was the source of early debate and confusion.  It was
 clarified in our later 2005 PRL work with Forbes et al \cite{Forbes:2004cr}. 
 
 Another type is of momentum structure falling off as a Gaussian\cite{Forbes:2004cr}
 \begin{equation}
   V^{II}_{\Bp\Bp^\prime}(\Bq) = g e^{-q^2/\lambda^2}\,, \qquad \forall \Bp,\Bp^\prime\,.
 \end{equation}
 Similar to type I interaction in a qualitative sense, for the BP phase to
 prevail, the falling off momentum scale $\lambda$ needs to be within the Fermi
 momentum difference between the two spins.
 
The BP state is similar to the BCS in terms of symmetry of the superfluid
order parameter---Cooper pairs condensed at zero momentum and the state is
uniform in position space.  Then, how does the state accommodate the spin
population difference $N_\up< N_\down$?   This question can be answered precisely by contrasting 
the many-body wavefunctions of the well-known BCS state
% BCS vs. Breached Pairing wavefunctions:
\begin{equation}
|\Psi_{BCS}\> = \prod_\Bp (u_\Bp + v_\Bp \psi^\dag_{\Bp\up}
\psi^\dag_{ -\Bp\down }) |0\> \,,
\end{equation}
and the new BP state,
\begin{equation}
|\Psi_{BP}\> =\prod_{\Bp: p<p^-_\Del} (u_\Bp + v_\Bp \psi^\dag_{\Bp \up}
\psi^\dag_{-\Bp \down}) \prod_{\Bp: p\in [p^-_\Del, p^+_\Del]}
\psi^\dag_{\Bp \down} 
\prod_{\Bp: p>p^+_\Del} (u_\Bp + v_\Bp \psi^\dag_{\Bp \up}
\psi^\dag_{-\Bp \down})
 |0\> \,.
\end{equation}
\paragraph{\bf Momentum-space phase separation.} The BP state accomplishes the
population imbalance by a mechanism that I would refer to as momentum-space
phase separation of superfluid and normal components.  The momentum region of
${\Bp: p\in [p^-_\Del, p^+_\Del]}$ is a normal component breach: it is filled by
only one fermion species, namely, the majority species (spin $\down$ in our
example here), hence there is no pairing. Superfluid components exist, in a way
similar to what happens in BCS state, in the momentum regions outside
this ``normal'' breach.

\begin{figure}[htp]
\begin{center}
\begin{tabular}{ll}
BCS&\\
\includegraphics[width=.6\linewidth]{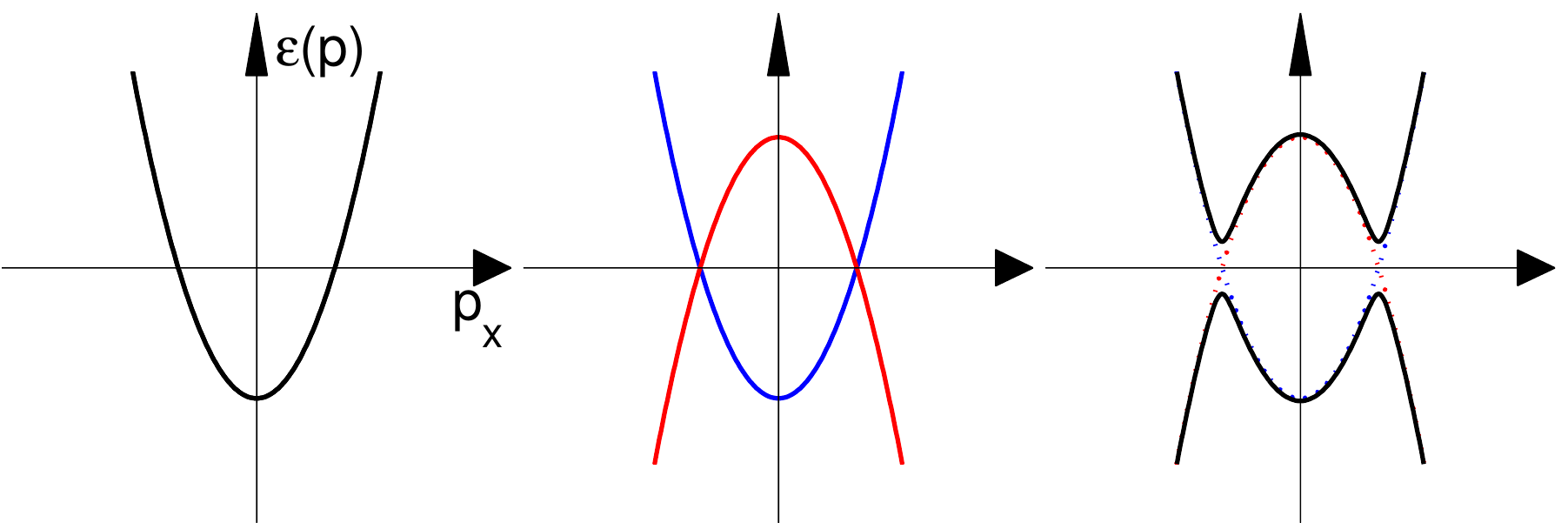} &
\includegraphics[width=.25\linewidth]{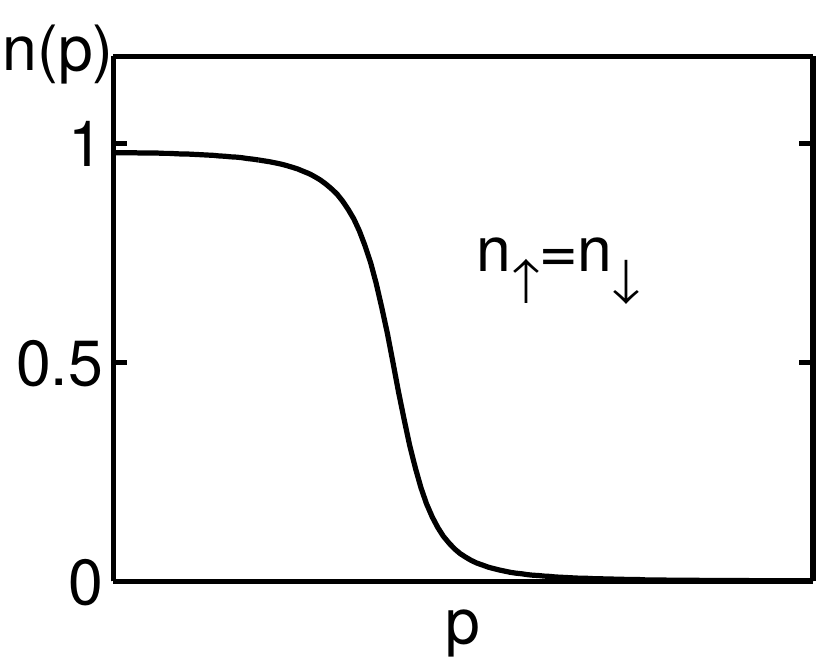}\\
  %\raisebox{-1.5in}[1ex][1ex]
  {BP}& \\
\includegraphics[width=.6\linewidth]{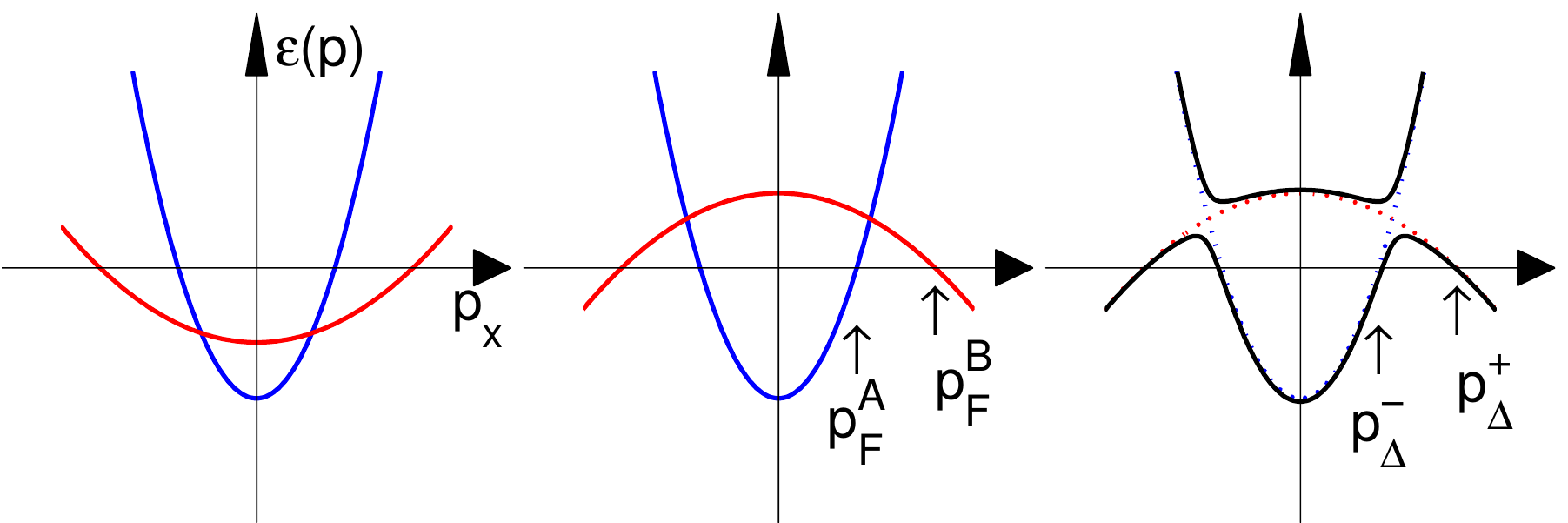} &
\includegraphics[width=.25\linewidth]{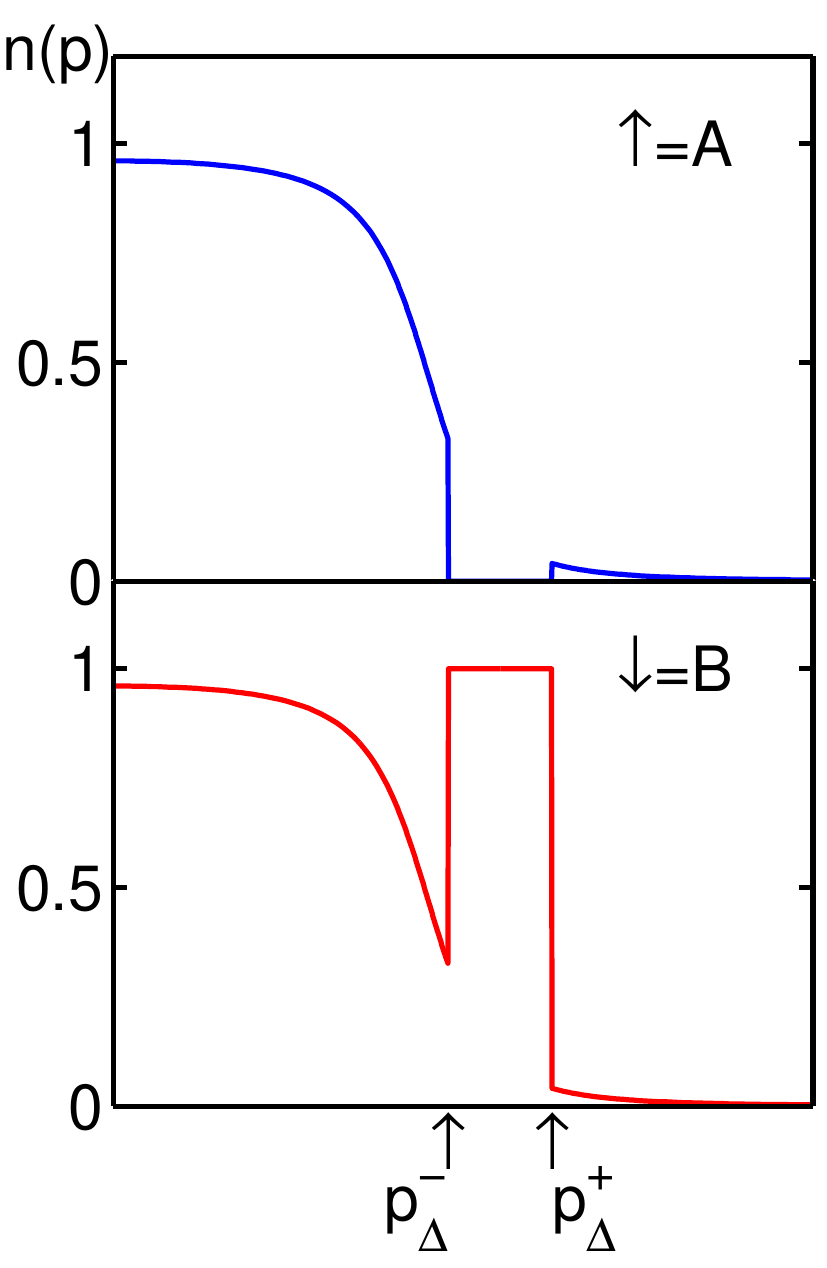}
\end{tabular}
\end{center}
\caption{Comparison of BP and BCS states revealed from excitation spectra and
  fermion occupation numbers. In this illustration of example, we assume the
  condition $m_\up \ll m_\down$ and $N_\up< N_\down$ with two species fermions
  denoted as $A,B$ (or $ \up,\down$ exchangeably). The region bounded between
  $p^\pm_\Del $ is a breach in the momentum space, where only the majority
  species is present and no pairing takes place.  Within the breach, the
  occupation numbers are 1 and 0 for the majority and minority fermion,
  respectively (right panel).} \label{fig:BCS-BP}
\end{figure}

The BP variational wavefunction consists of the usual coherence variables
$(u_\Bp, v_\Bp)$ as well as additional variational parameters $p_\Del^{\pm}$, to
be determined by variationally minimizing the ground state energy with fixed
chemical potentials and solving the energy gap equation self-consistently.  The
following summarizes the results obtained in this manner,
\begin{equation}
\left\{
\begin{array}{c}
|u_\Bp|^2 \\ |v_\Bp|^2
\end{array}\right\}
 =  {1\over 2} \left( 1 \pm {\eps^+_\Bp \over
\sqrt{{\eps_\Bp^+}^2 + |\Del|^2 }}\right) \,, \qquad
\eps^\pm_\Bp \equiv \frac{\eps_\up(\Bp)\pm \eps_\down(\Bp)}{2}
\end{equation}
with $ \epsilon_\sig(\Bp) = \Bp^2/2m_\sig - \mu_\sig$.  
The Bogoliubov quasi-particle spectrum in the BP superconducting state takes a
different form than in the BCS state, shown as follows: 
\be
E^\pm_\Bp = \eps^-_\Bp \pm \sqrt{{\eps_\Bp^+}^2 +  |\Del|^2 }\,.
\ee
Taking the usual notation,
$\Del$ denotes the excitation energy gap appearing in the superconducting state.
What is special in the BP excitation spectrum is the presence of gapless Fermi
surface while being superconducting (or superfluid for charge neutral
fermions).   The gapless surfaces are the zero mode solution of the
quasi-particle excitation in momentum space, 
\begin{equation}
  E^+_\Bp E^-_\Bp =0\qquad  \Rightarrow  \qquad p=p^\pm_\Del \,.
\end{equation}
The peculiar features of the BP state are illustrated in Fig.~\ref{fig:BCS-BP}.
The momentum dependence of the fermion occupation numbers, $n_\sig(\Bp)$, tells
two important points. First, it shows where the pairing spectral weight is
taking place most. That is the region where $n_\sig(\Bp)$ deviates from $1$
most due to fermion pairing. Note this is the ground state property ($T=0$) under discussion. In
the example we adopt here ($m_\up\ll m_\down$ and $p_F^\up < p_F^\down$), Cooper
pairing occurs mostly around the smaller Fermi surface, hence initially called
``interior gap'' superfluidity\cite{Liu-Wilczek:03}. If one had considered the opposite case (still
$m_\up\ll m_\down$ but $p_F^\up > p_F^\down$), one would find fermion pairing
should occur most along the larger Fermi surface, and then one would see
something like ``exterior gap''\cite{Gubankova+:03}.   Second, the occupation numbers tell where the
breach (a region of no Cooper pairing) is precisely located. It is the place in
momentum where the
minority species is 0 and the majority is 1.  In both ``interior'' and
``exterior'' gap case, the breach is the hallmark of the new
superfluid.\cite{Gubankova+:03,Forbes:2004cr}

Let us summarize the characteristic features of the BP phase: 
\begin{enumerate}[(a)]
\item It realizes a momentum-space phase separation to accommodate the population
imbalance between two spins.
\item It exhibits coexisting superfluid and normal Fermi liquid components in
the quantum ground state through such a momentum-space phase separation.
\item It is a superfluid with a full surface of gapless quasiparticle
  excitations. In the case illustrated in Fig.~\ref{fig:BCS-BP}, the BP phase
  has 2 gapless surfaces (BP2). At certain critical point, one gapless surface is
  found possible
  (BP1).\cite{Son-Stephanov:06,Yang-Sachdev:06,Sachdev-Yang:06,Liu+Zoller:04}
\item Unlike the fully gapped BCS phase, the superfluid density is not equal to
the total fermion density, but to the density of only those fermions filled outside the breach. 
\item Unlike the FFLO phase, it does not spontaneously break the translational and rotational
  symmetries. 
\end{enumerate}

\section{Remarks} 
Since the initial proposal, the concept of interior gap and BP phase has
received much attention and debates regarding its stability and competition with
other viable
phases\cite{Wu-Yip:03,Pao-Wu-Yip:06,Sheehy+:prl:06,Bulgac+Forbes:06}. The
confusion has something to do with that the BP phase we proposed is related to
one of the solutions to the superconducting gap equation studied by
Sarma\cite{Sarma:63} in early years, which was correctly realized as a unstable
solution at his analysis.  (It is sometime jointly called Sarma or BP phase in
literature). The key point is that we introduced the BP phase, despite some
similarity to Sarma's state in the structure of wavefunction, with a large mass
ratio. This is one of the key ingredients to have a physically stable state, to
be highlighted next.

Based on studies and discussions so far, our general conclusion is that the BP
phase requires two crucial conditions\cite{Forbes:2004cr}: (a) a relatively
large mass ratio between the two fermion species and (b) strong momentum
dependence of a two-body attractive $s$-wave interaction. The latter is
equivalent to requiring strong spatial dependence of the interaction in position
space, by Fourier transformation.  The examples we have found include the
Gaussian type of long range interaction in position space or the type of
interaction restricted to a narrow momentum strip as in the BCS model. The BP
phase has been found as a competing phase in the phase diagram by other studies.
An effective field theory approach by Son and Stephanov \cite{Son-Stephanov:06}
showed a universal phase diagram of a homogenous gapless superfluid phase which
corresponds to the BP phase.  The quantum critical theory based on the
renormalization group analysis by Yang and Sachdev
\cite{Yang-Sachdev:06,Sachdev-Yang:06} showed the gapless BP phase is stable in
a 2D superfluid.  Dynamical mean field theory by Dao et al \cite{Dao+Georges:08}
found that the attractive Hubbard model yields a polarized phase closely
connected to the physics of the Sarma or BP phase (with two fermi surfaces) down
to very low temperatures.  Furthermore, Dukelsky et al obtained
the phase diagram of an exact solvable model by using the algebraic
Richardson-Gaudin techniques, which shows not only a breached pair (BP) phase
but also another exotic phase that they dubbed ``breached'' LOFF
phase~\cite{Dukelsky-Ortiz:06}.  This model generalizes what was introduced by
me with Frank \cite{Liu-Wilczek:03}.

The presence of gapless Fermi surfaces was found to manifest itself by some
remarkable, unconventional aspects such as inducing a spatially oscillating
potential between superfluid vortices, akin to the RKKY indirect-exchange
interaction in non-magnetic metals \cite{Stojanovic:08}.  It was remarked that
the ``interior gap''/BP superfluidity might be relevant to explain the
experimental observation of unpaired electrons present in the heavy-fermion
superconductor CeCoIn$_5$\cite{Taillefer:05}.  In the context of high density
QCD, gapless color superconductivity---a phase related to the Sarma/BP---was
proposed as a natural candidate for quark matter in the cores of compact stars
at zero and finite temperature by Huang and Shovkovy through a detailed analysis
\cite{Huang-Shovkovy:03} and their earlier paper \cite{Shovkovy-Huang:03}.
% For
% trapped cold ensembles of atoms, experimental signatures detecting breached pair
% phases with 1 or 2 Fermi surfaces (BP1 and BP2) were put forward by Yi et al
% \cite{Yi-Duan:06,Lin-Yi-Duan:06} through their systematically analyzing the
% effect of mass and population imbalance.
For trapped cold ensembles of atoms, the effect of mass and population imbalance
in gapless superfluidity was analyzed by Bedaque, Caldas, and Rupak
\cite{Bedaque:03,Caldas:04} (where the authors did not consider
momentum-dependent interaction) suggesting real-space phase separation of normal
and superfluid components and by Yi and Duan and later by Lin with them
\cite{Yi-Duan:06,Lin-Yi-Duan:06} putting forward experimental signatures
detecting breached pair phases with 1 or 2 Fermi surfaces (BP1 and BP2).

% For
% trapped cold ensembles of atoms, understanding the effect of
% mass and/or population imbalance was advanced by an outburst of systematic studies
% and predictions  of peculiar phenomena including the phase-separation of normal
% and superfluid components in real
% space~\cite{Bedaque:03,Caldas:04,Sheehy+:prl:06,Pao+Wu+Yip:06} and 1- or
% 2-Fermi surface breached phases with experimental detecting signatures
% \cite{Yi-Duan:06,Yi-Duan:06pra}, 

\section*{Acknowledgement}
I am grateful to M. Forbes, E. Gubankova, M. Huang, Y. B. Kim, G. Ortiz,
K. Rajagopal, L. Taillefer, F. Wilczek, and K. Yang for valuable discussions. This work is
supported by AFOSR Grant No. FA9550-16-1-0006, the MURI-ARO Grant
No. W911NF17-1-0323 through UC Santa Barbara, and the Shanghai Municipal Science
and Technology Major Project (Grant No. 2019SHZDZX01).

%% I am grateful for xxxx

\bibliographystyle{ws-rv-van}
\bibliography{fermi-pair,liu-papers2020c}
\end{document}